\documentclass[journal=jprobs,manuscript=article,etalmode=firstonly]{achemso}
\setkeys{acs}{etalmode=firstonly, maxauthors=4}
\usepackage{soul}
\usepackage{color,xcolor}
\usepackage[hidelinks]{hyperref}
\usepackage{chemformula} 
\usepackage[T1]{fontenc} 
\usepackage{subcaption}
\usepackage{booktabs, array, tabularx, threeparttable}
\usepackage{multirow}
\usepackage{achemso}
\usepackage{clrscode3e,url}
\usepackage[USenglish]{babel}
\usepackage[useregional]{datetime2}

\newcommand{\mydate}[3]{\DTMdisplaydate{#1}{#2}{#3}{-1}}

\hypersetup{
    colorlinks,
    linkcolor={red!90!red},
    citecolor={blue!50!black},
    urlcolor={cyan!100!black}
}

\graphicspath{{Final_Figures/}}
\author{$\text{Vikram Singh}^{\dagger}$}
\affiliation{Centre for Computational Biology and Bioinformatics, Central 
University of Himahcal Pradesh, Dharamshala, India}
\author{Vikram Singh}
\affiliation{Centre for Computational Biology and Bioinformatics, Central 
University of Himahcal Pradesh, Dharamshala, India}
\email{vikramsingh@cuhimachal.ac.in}

\title{C19-TraNet: an empirical, global index-case transmission network of SARS-CoV-2}


\begin{document} \selectlanguage{USenglish}

\begin{abstract}

Originating in Wuhan, the novel coronavirus, severe acute respiratory syndrome 2 (SARS-CoV-2), has astonished health-care systems across globe due to its rapid and simultaneous
spread to the neighboring and distantly located countries. To gain the systems level understanding of the role of global transmission routes in the COVID-19 spread, in this study, we have developed the first, empirical, global, index-case transmission network of SARS-CoV-2 termed as C19-TraNet. We manually curated the travel history of country wise index-cases using government press releases, their official social media handles and online news reports to construct this C19-TraNet that is a spatio-temporal, sparse, growing network comprising of $187$ nodes and $199$ edges and follows a power-law degree distribution. To model the growing C19-TraNet, a novel stochastic scale free (SSF) algorithm is proposed that accounts for stochastic addition of both nodes as well as edges at each time step. A peculiar connectivity pattern in C19-TraNet is observed, characterized by a fourth degree polynomial growth curve, that significantly diverges from the average random connectivity pattern obtained from an ensemble of its $1,000$ SSF realizations. Partitioning the C19-TraNet, using edge betweenness, it is found that most of the large communities are comprised of a heterogeneous mixture of countries belonging to different world regions suggesting that there are no spatial constraints on the spread of disease. This work characterizes the superspreaders that have very quickly transported the virus, through multiple transmission routes, to long range geographical locations alongwith their local neighborhoods.

\end{abstract}

\section{Introduction}

Recent outbreak of COVID-19 has been emerged due to a new species of human 
infecting coronaviruses (HCovs), namely, SARS-CoV-2 \citep{of2020species} that 
was isolated 
from a cluster of pneumonia cases in Wuhan, capital city of Hubei Province, 
China \citep{zhu2020novel} in December 2019. It rapidly spread across the globe 
and was 
declared as a Public Health Emergency of International Concern (PHEIC) on 
\mydate{2020}{1}{30} \citep{world2005statement}, and was later declared a 
pandemic on \mydate{2020}{3}{11} by the World Health Organization (WHO) 
\citep{world2020director}. As of \mydate{2020}{6}{25}, more than $9.1$ million 
confirmed cases, and more than $0.47$ million deaths have been reported due to 
COVID-19 in $216$ countries/territories \citep{WHOSR109}. The outbreak is 
attributed to a zoonotic transmission of SARS-CoV-2 from primordial or 
intermediate host to humans, it could have further propagated through 
human-to-human contacts \cite{wang2020clinical,bai2020presumed}. Globalization 
of transport systems \textit{via} air, sea and land routes has resulted in 
drastic decrease in journey time alongwith an unprecedented increase in 
transport volume of passengers. Although global transport has played a pivotal 
role in world's economic growth, it has also facilitated the disease causing 
agents to move to remote and otherwise unreachable places very quickly 
\citep{tatem2006global}, and therefore has enhanced the potential of any endemic 
to become a pandemic \citep{fidler1996globalization}.

In the past five centuries the world has witnessed several major pandemics, 
like, plague \citep{keeling2000metapopulation} (caused 
by \textit{Yersinia pestis}), cholera \citep{sack2004sack} (\textit{Vibrio 
cholera}), influenza \cite{ferguson2003ecological} (influenza virus), HIV/AIDS 
\citep{perrin2003travel} (HIV), SARS \citep{peiris2004severe} (SARS-CoV) etc., 
that had badly affected most of the world population as well as economy. All 
these pandemics had been originated in localized regions and had different means 
of 
human-to-human transmission, however, they were transported to distant 
geographical locations by human movements \textit{via} 
air, land or sea transport \cite{zinsser1935rats, reidl2002vibrio, 
palese2004influenza, 
heymann2004international}. There are many other emerging, reemerging 
infections, with varying consequences in terms of severity, morbidity, 
mortality, which the world continues to confront \cite{bloom2019infectious}. 
Different epidemiological strategies, like, modeling disease dynamics, spread, 
mitigation and other epidemiological parameters of different outbreaks have 
been applied successfully in the past 
\cite{epstein2009modelling,gates2018innovation,weiss2003hiv}. Network theoretic 
approaches have also been successfully utilized to gain understanding about the 
topology of different 
epidemiological disease transmission networks 
\cite{salathe2010high,enright2018epidemics,keeling2005networks,
craft2015infectious}. Topological understanding of infection transmission 
networks provides us critical insights about infection growth and distribution. 
Furthermore, knowledge of potential transmission routes allows us to devise 
effective control strategies to contain infection \cite{bell1999centrality}.

Due to lack of specific medication for SARS-CoV-2, non-pharmaceutial 
containment based intervention methods have become 
indispensable tools to grapple with the infection. A large number of studies 
are being published to investigate important metrics, like, CFR, basic 
reproduction rate ($R_0$), spread kinetics etc., however, a comprehensive study 
investigating the transmission routes of the virus across countries is still 
lacking. The actual structure (topology) underlying the global spread of the 
SARS-CoV-2 is also less explored. To prevent virus re-invasion in the 
countries/territories/regions (from where the virus has been eradicated) 
considering large incubation period of SARS-CoV-2, at this stage of outbreak, it 
is crucial to gain insights about the short range human-to-human contacts as 
well as long range geographically separated country-to-country global 
transmission pathways. It may have implications in making SARS-CoV-2 
prevention policies as well as better prepare ourselves to fight against 
future outbreaks. In this study, we have constructed an empirical global 
transmission 
network of SARS-CoV-2 and explored its topological space to explore the 
mechanism of global spread of infection. Increased topological knowledge will 
provide us the understanding about the epidemiological dynamics of the virus 
invading remote geographical locations. 

\section{Methods}

\subsection{Network construction}
In this work, we have constructed the index-case empirical global transmission 
network of SARS-CoV-2 by manually going through the daily situation reports 
published by WHO, government press releases of various countries, their 
official social media handles (facebook, twitter), and other online news 
reports providing relevant information regarding the travel history of the 
patients for a period of $84$ days ranging from \mydate{2020}{1}{13} to 
\mydate{2020}{4}{5}. The transmission network is a dynamic network having the 
potential to grow on daily basis. Countries comprise the nodes of this network, 
while infected index-case individuals carrying the infection from a given 
country $X$ to some other country $Y$ connect these two countries with a 
directed link with $X$ as the source node and $Y$ as the target node. The 
direction of an edge indicate the flow of virus infection across borders. The 
connection between two countries have been placed on the basis of travel 
history of the patient. Any foreign national reported to be corona positive in 
any country has been considered as a carrier of virus to that country 
irrespective of his home nation. For the countries reporting more than one 
patient as their index-cases, if the reported patients  had different travel 
histories we first observed the timeline to find the countries in which virus 
had been already spread out and if the the virus had already reached in the 
source countries then we placed a link between all these source countries with 
the target country. On the other hand for the source countries where the 
virus had yet not reached, we considered that the patient has contracted the 
virus \textit{enroute} and did not place link between them.

\subsection{Stochastic scale free (SSF) random network models}

SARS-CoV-2 or any other outbreak that spread through human-to-human 
contacts will follow a peculiar spread pattern where the infection starts from 
a source area and then gradually spreads across other countries/areas/regions. 
To model this specific spread pattern, we are proposing a novel algorithm, 
namely, stochastic scale free (SSF). SSF is developed as an extension of the 
basic Barab\'{a}si-Alberts (BA) model of scale free network. In the proposed 
SSF algorithm, network grows stochastically over time with number of nodes 
and their associated edges both being drawn from an appropriate distribution. 
The algorithm follows three basic steps: (\romannumeral 1) The 
infection spread is unidirectional \textit{i.e.} from source to target, new 
nodes (targets) attach preferentially with already existing nodes (source) and 
have linkage probabilities proportional to source degree. (\romannumeral 2) 
At every time step, any number of nodes ($0, 1, 2, \dots,\id{maximum-nodes}$) 
can be incorporated into the network. (\romannumeral 3) Every node can bring 
one or more number of links ($1, 2, 3, \dots, \id{maximum-nodes}$) with it. The 
key purpose of stochastically adding both nodes and edges is to model the 
index-case of SARS-CoV-2 infection across various countries as a growing 
transmission network. In this network, the index-case import in various 
countries may have two different scenarios, namely, (i) more than one country 
can import their index-case on any given day \textit{i.e.} more than one nodes 
may be added at a single time step. (ii) a country can receive their index-case 
from more than one countries \textit{i.e.} an incoming node may have more than 
one in-degree.  

The main characteristic of the growth of C19-TraNet is that initially very few 
new countries get infection so the probability of infection transmission 
to new countries at any given day is very low, however, this probability 
increases very rapidly with increase in number of infected countries. 
Therefore, while generating an SSF model of C19-TraNet, to obtain the number of 
nodes and links associated with them at a unit time interval $t$ (one day), we 
have used beta distribution that is given by $f(x;\alpha,\beta) 
= \frac{\Gamma (\alpha + \beta) }{\Gamma (\alpha)  \Gamma(\beta)} x^{\alpha - 
1}(1 - x)^{\beta-1}$ where $\Gamma$ is a gamma function, and $\alpha$, $\beta$ 
are two non-zero parameters that control the shape of this distribution. The 
scaling parameters at any time interval $t$ were computed as $\beta_t = 
\frac{\id{Max-time} - \id{Current-time} }{\id{Max-time}}$ and $\alpha_t = 1 - 
\beta_t$, where $\id{Max-time}$ is the number of days for which the infection 
is to be modeled ($84$ in the current study) and $\id{Current-time}$ is the 
$t^{th}$ time step. On the other hand, scaling parameters $\alpha = 0.05$ and 
$\beta = 0.95$ were qualitatively selected to draw the number of edges 
asociated with the incoming nodes. The small value for $\alpha$ and a high 
value for $\beta$ are  deliberately selected considering a very low average 
connectivity ($\approx 2.1$) of C19-TraNet. The maximum number of nodes 
($\id{maximum-nodes}$) that can be incorporated into the network at a discrete 
time step is given by $\id{maximum-nodes} = \langle k_{in + out}  \rangle + 1 
\sigma $, while maximum number of edges that a node can bring along with it was 
computed as $\id{maximum-nodes} = \langle k_{out} \rangle + 1 \sigma $. Values 
obtained were truncated to retain only the integer part. Corresponding to the 
manualy curated C19-TraNet, we constructed an ensemble 
of $1,000$ stochastic scale free random networks having same number of 
nodes, and approximately same number of edges.

\subsection{Global network metrics and community detection}
Four global network metrics, namely, (i) degree ($k$) 
distribution, (ii) network density $(\rho)$, (iii) average clustering 
coefficient $(C)$, and (iv) average path length $(L)$ were estimated for 
C19-TraNet and compared with the corresponding $1,000$ SSF random network 
models generated \textit{via} proposed SSF algorithm. The degree of a node is 
the number of edges connected with it. In a directed network, the number of 
edges incident on a particular node constitute its indegree ($k_{in}$), while 
the number of edges extending outward constitute its oudegree ($k_{out}$) 
\citep{Newman2003}. The in and out degree distributions of a network are 
defined as fraction of nodes having $k$ incoming or outgoing links. 
Furthermore, we computed the density of network that is defined as a fraction 
of existing edges in the network to the all possible edges 
\textit{i.e.} $\rho = \frac{m}{ \binom{n}{2}}$. We also enumerated clustering 
coefficient for every node that is defined as the number of edges existing 
between first neighbors of the node $i$ to the total possible edges existing 
between the first neighbors, averaged over all the nodes $i$ 
\citep{watts1998collective}. If $N_i$ is the set of nodes directly connected to 
node $i$ then its clustering coefficient for a directed network is defined as 
$$C_i = \frac{|\{e_{jk}: v_j, v_  \in N_i, e_{jk} \in E\}|}{|N_i| (|N_i| - 1) 
}$$, where  $E$ is the edge set of network $G(V,E)$. Average path length of a 
network is the average of geodesic paths between all possible pairs of nodes in 
the network \citep{watts1998collective} and is defined as $$L = 
\frac{2}{n(n-1)} \sum \limits_{a = 1}^{n-1} \sum \limits_{b = a + 1}^{n} 
\delta_{ab}$$. We identified communities based on edge betweenness, where the 
network is 
partitioned into node sets by removing high betweenness edges 
\citep{newman2004finding}. We generated $1,000$ random realizations of SSF 
models and Mann-Whitney U test was performed on distributions of 
average outdegree obtained at different time steps for both real and random 
networks \citep{mann1947test}.

\section{Results and Discussion}
\begin{figure}
     \centering
        
     \begin{subfigure}[b]{1\textwidth}
         \includegraphics[width=\textwidth]{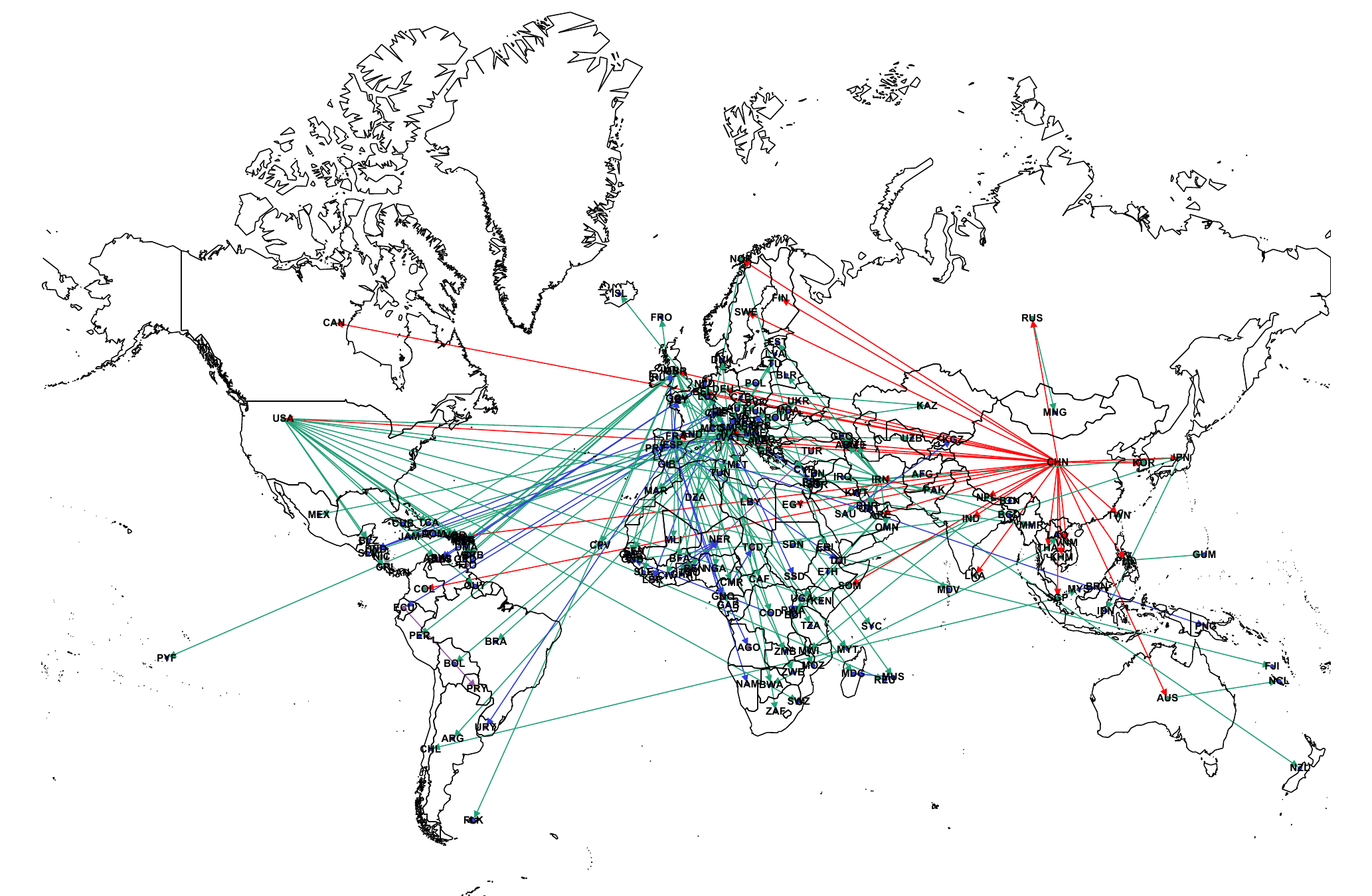}
         \caption{}
         \label{fig:Net}
     \end{subfigure}
     ~
     \begin{subfigure}[b]{0.43\textwidth}
         \includegraphics[width=\textwidth]{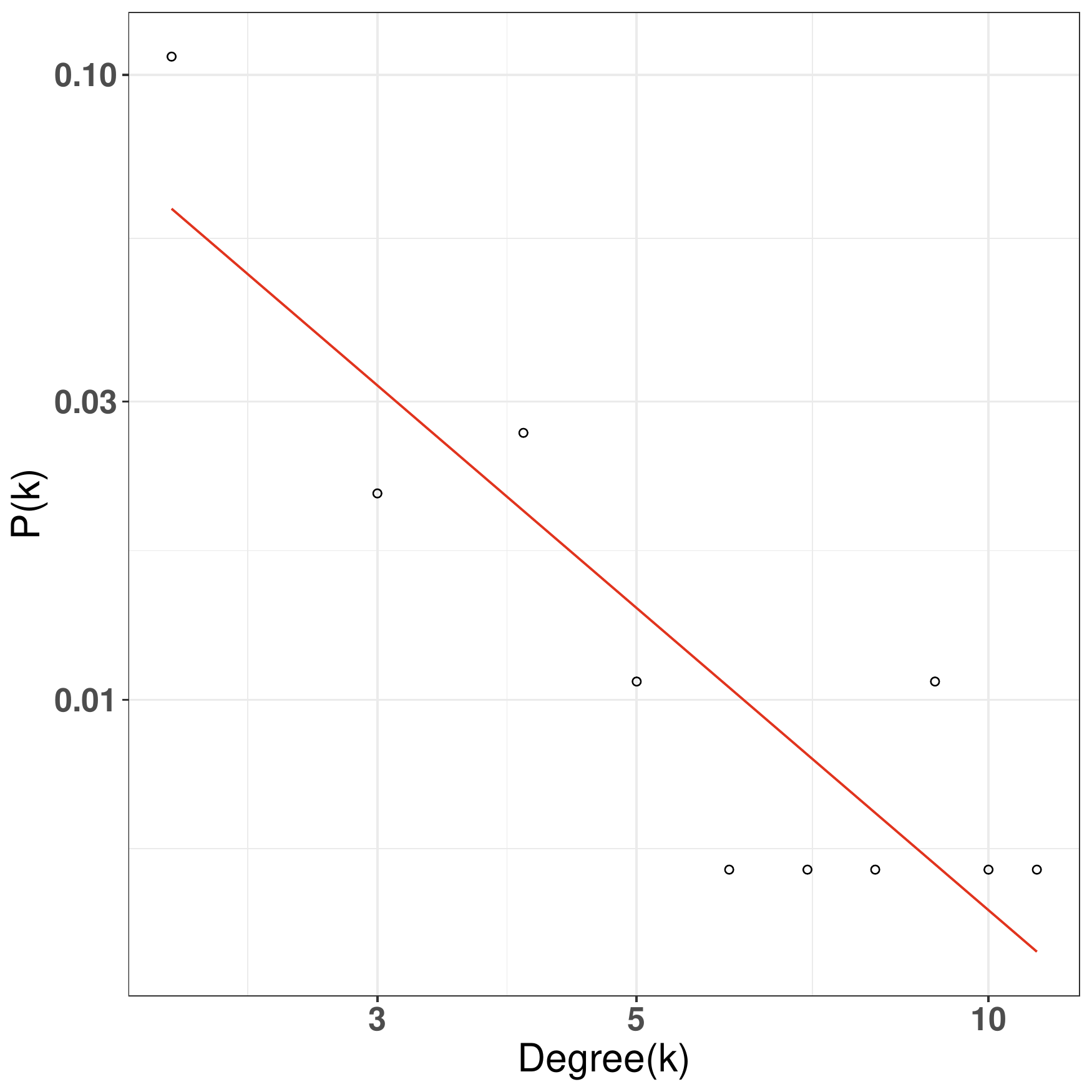}
         \caption{}
         \label{fig:DDist}
     \end{subfigure}

     \caption{(a) Empirical global COVID-19 transmission network (C19-TraNet) 
with edges highlighted as per the distance of source nodes from the country where the first COVID-19 patient was reported, into three colors: red (primary 
spreader, \textit{i.e.} China), green (secondary spreaders)  and blue (tertiary spreaders). (b) Degree distribution of C19-TraNet with a power law fit having exponent $-1.63$.}
\end{figure}

The dynamic effects of heterogeneous communication routes has been central 
to the propagation of many diseases \cite{Mangili2005}, hence the interplay 
between them and the underlying transmission network of SARS-CoV-2 across 
different countries/territories/areas is the focus of this study. We catalogued 
the first person diagnosed to have contracted SARS-CoV-2 virus for $187$ 
countries. By investigating the travel history of index-cases from government 
press releases, their official social media handles and other online news 
reports, we inferred a probable source (country where the patient contracted the 
virus) and the target (country confirming COVID-19 infection in the diseased) 
of each edge to be incorporated in the empirical COVID-19 global transmission 
network (C19-TraNet). The constructed C19-TraNet is comprised of $187$ nodes 
and $199$ directed edges specifying the route of infection transmission from 
the source country to the target country (\ref{fig:Net}). The final network is 
a combined snapshot representing SARS-CoV-2 transmission 
across the countries that has grown over a period of $84$ days from 
\mydate{2020}{1}{13} to \mydate{2020}{4}{5}.

\subsection{SARS-CoV-2 global transmission routes}

The outbreak originated from China (Western Pacific region) 
\mydate{2019}{12}{30} and within two months (by \mydate{2020}{02}{25}) the 
virus had been spread to at least one country of all the six WHO regions. By 
the end of January, the contagion had been reported in five WHO regions, 
namely, Western Pacific Region ($45 \%$), South-East Asia Region ($40 \%$), 
European Region ($13.33$), Region of Americas ($5.56 \%$) and Eastern 
Mediterranean ($4.54 \%$). The numbers in parentheses indicate 
percentage of infected countries per region. Except Malaysia and Spain, which 
imported virus from Singapore and Germany, respectively, all the other $23$ 
countries imported it from China. Global transmission of SARS-CoV-2 can broadly 
be attributed to two tidal waves, namely, (\romannumeral 1) a Chinese wave, 
from the date of origin of outbreak till the end of January, 2020 and 
(\romannumeral 2) a European wave that took off from mid February and continued 
until almost all the countries got infection.  The major reason for this quick 
spread is, (\romannumeral 1) emergence of China as an economic globaliser in 
past few decades that has increased international business and trade, and 
(\romannumeral 2) expansion of highly connected local and global travel systems 
\cite{trindade2018globaliser}. 

Travel data suggests that approximately $3$ billion people traveled through 
trains or flights during the Chunyun (Lunar New Year) holidays 
\cite{daszak2020strategy}. Also, a large volume of passengers have been 
estimated to travel within \cite{zhao2020association} and outside China 
\cite{bogoch2020pneumonia} through land and air routes from previous travel 
data. These studies suggest a significant relation between transport volume and 
likelihood of imported cases of SARS-CoV-2. Although ban was implemented in 
Wuhan city on \mydate{2020}{1}{23}, which was followed by travel restrictions 
to $14$ of its neighboring cities of Hubei Province on \mydate{2020}{1}{24}, 
however, by this time the virus had been transmitted to other Chinese cities as 
well as to $9$ different countries belonging to four WHO regions \cite{WHOSR5}. 
From these sources, the virus kept on spreading locally as well as outside 
China till the end of January when the results of containment measures became 
visible with overall decrease in cases exported from China 
\cite{chinazzi2020effect} (diminishing the effects of Chinese wave). Then from 
mid of February, second tidal wave took off from European Region and 
circulated the infection to most parts of the globe. Italy, Iran and Spain were 
among the super spreaders which aggressively spread the contagion. Owing to 
large incubation period of SARS-CoV-2 along with lack of necessary 
infrastructure, sophisticated testing facilities, herd immunity, a large number 
of undocumented asymptomatic or presymptomatic cases passed undetected into 
susceptible populations of  different countries/territories/areas, thus 
provided unprecedented routes to virus transmission 
\cite{li2020substantial,rothe2020transmission}. These large 
number of possible routes also increased the strength of second wave by 
providing opportunities to circulate the virus to most of the uninfected 
countries belonging to different regions. Therefore, we may infer that a large 
incubation period, high movement and multiple routes helped global circulation 
of the disease.

\subsection{Global topological metrics of C19-TraNet}

The final network is sparse with a very low density ($0.01$). This is expected, 
as we constructed only the index-case transmission network. Average 
connectivity of the network is found to be $2.128$, computed by ignoring the 
edge directions. Average or characteristic path length (APL) of a network 
represents closeness in the network, so lesser the magnitude of APL closer 
the nodes are. The characteristic path length of C19-TraNet is found to 
be $1.56$, while its diameter and radius are found to be $4$ and $1$, 
respectively. As per APL, most of the countries are infected by one or two 
transmission events. Network diameter suggests that most distant country in the 
network is infected by a maximum of four transmission events. These short path 
related metrics explain the quick global spread of virus, \textit{i.e.} the 
virus got access to those countries which have a wider reach on the globe that 
helped in propelling the infection across the world in a very short span of 
time. Short APL and a very small network diameter also indicate that there are 
some long range connections connecting distant geographical locations bringing 
an element of small worldness to the network. There are some hub nodes (very 
few) with a large number of edges in C19-TraNet. Plotting the
degree distribution on log-log scale, we obtained a straight line (with scaling 
exponent $-1.63$) indicating that C19-TraNet follows power law degree 
distribution and possess the property of scale freeness (Fig. \ref{fig:DDist}). 
The power law pattern of degree distribution has a simple interpretation that 
only few countries in the network are responsible for spreading the virus to 
most other countries. It has been reported that for any scale free disease 
transmission networks with scaling exponent less than $3$, the virus 
reproduction number ($R_0$) is greater than $1$ \textit{i.e.} very few nodes 
(super spreaders) are sufficient to maintain virus load in the susceptible 
populations \citep{schneeberger2004scale}. Its main implications lie in 
designing control strategies, like, targeting the super spreaders by 
implementing the basic containment strategies \textit{e.g.} prohibit transport, 
sanitization etc..

\subsection{Global infection exportation potential (GIEP)}

Emergence of a large number of countries in C19-TraNet as leaf nodes (having 
very low average clustering coefficient) implies that the infection was not 
equally spread by all the countries, rather only a few countries were 
responsible for its spread. We observed that only $38 (\approx 20 \%)$ 
countries among $187$ have at least one edge projecting outward. Among them, 
only seven countries are found to have an outdegree of $10$ or more, with Italy 
at the top of the tally contributing a total of $47$ exportation events 
followed by china ($29$) and Iran ($16$) (see Table \ref{tab:Prop}). 
Furthermore, infection import to every target node in C19-TraNet would have 
occurred in two possible ways, either a domicile of the target country has 
brought the virus from a source country, or a foreign national 
visiting a source country has spread the infection to the target country. To 
investigate the proportion of these two cases in the global spread of 
infection, we enlisted the nationalities of SARS-CoV-2 carriers and found that 
most of the index-cases are local nationals who contracted the infection in 
some source country and then brought it to their home nations. 

\begin{figure}
     \centering
        
     \begin{subfigure}[b]{0.55\textwidth}
         \includegraphics[width=\textwidth]{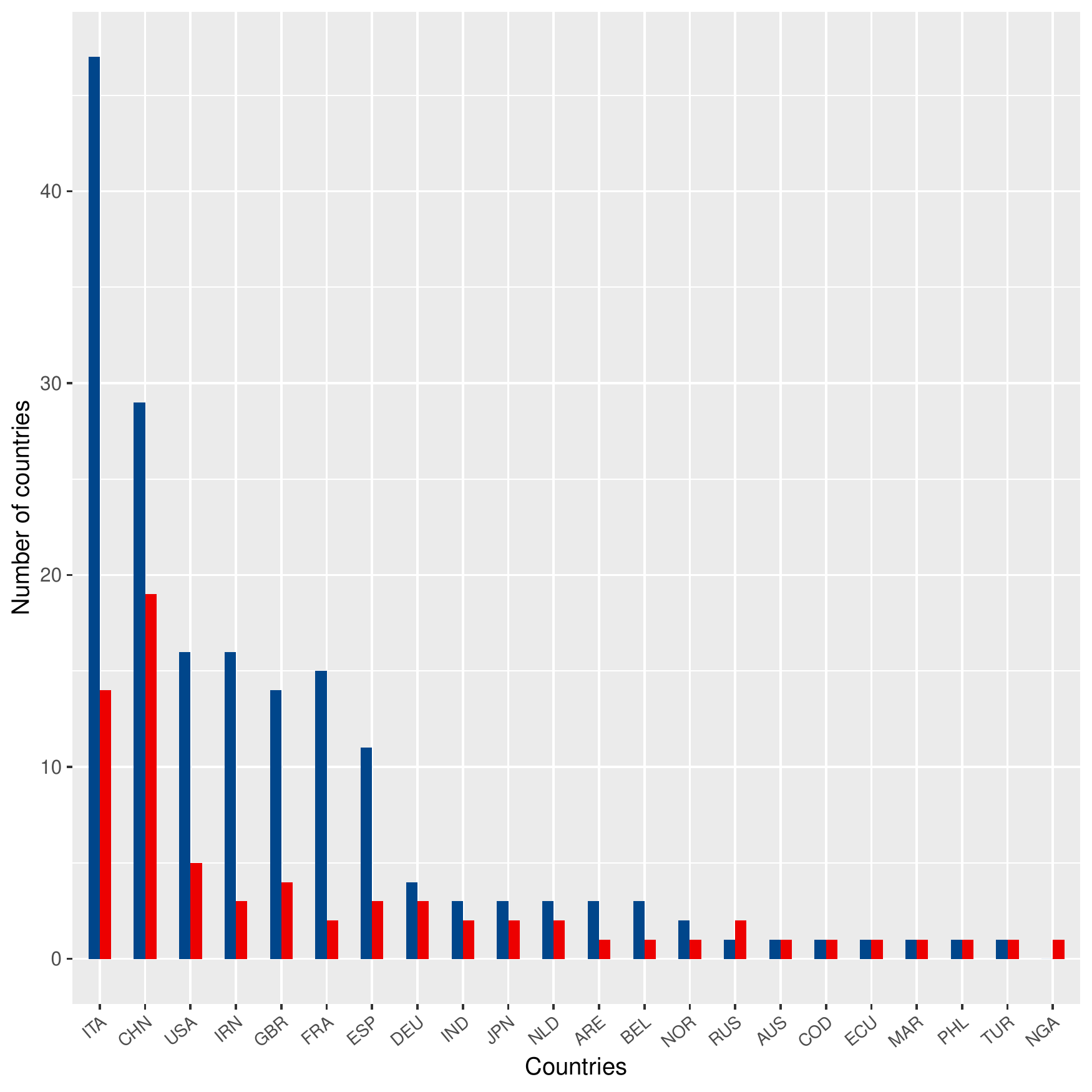}
         \caption{}
         \label{fig:OutDeg}
     \end{subfigure}
     ~
     \begin{subfigure}[b]{0.55\textwidth}
         \includegraphics[width=\textwidth]{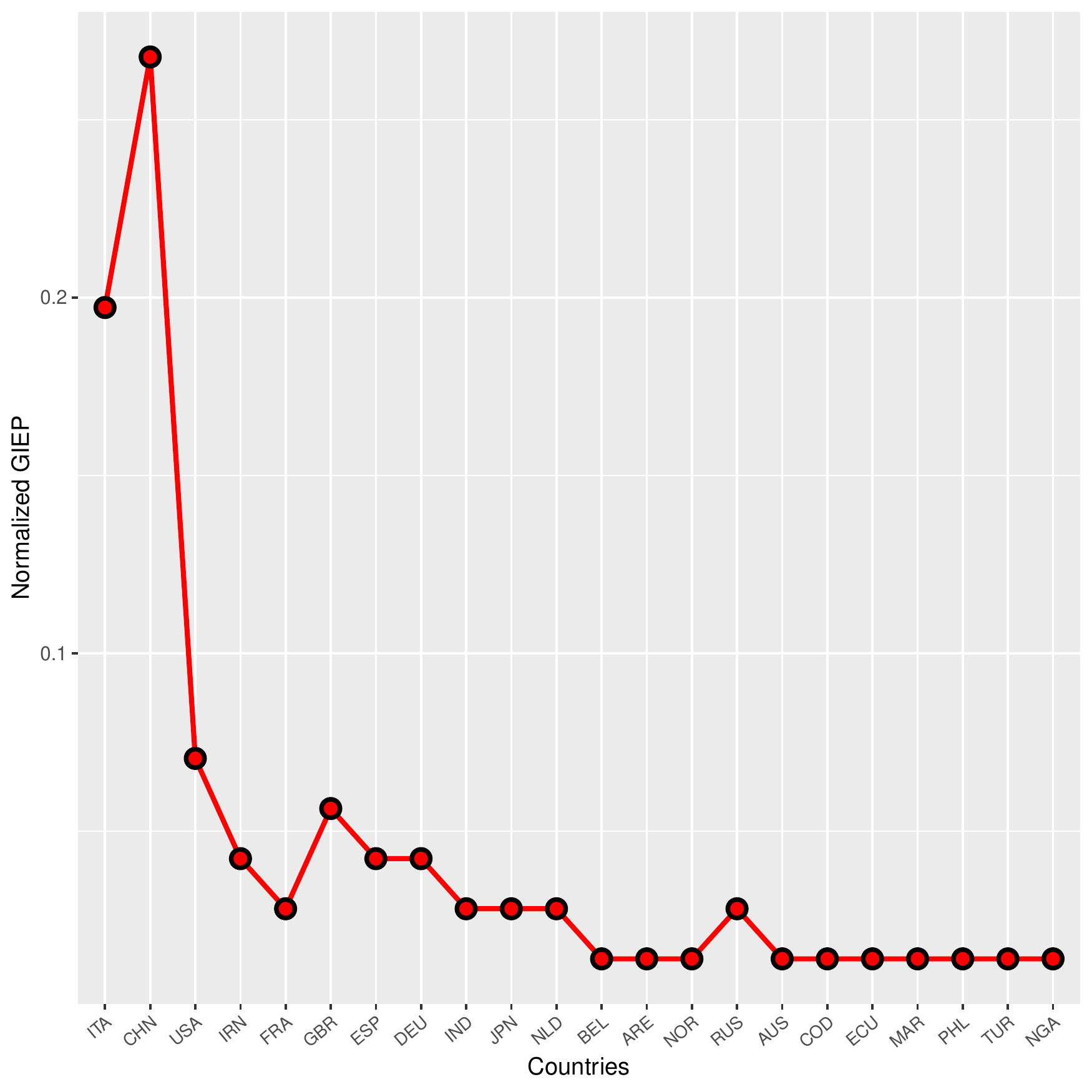}
         \caption{}
         \label{fig:ISP}
     \end{subfigure}

     \caption{ Outdegree histogram and global infection 
exportation potential (GIEP) of $22$ source countries: (a) Total number of 
index-case exportation events from a source country (blue), and the number of 
such events carried out by its domiciles (red). (b) Global infection 
exportation potential (GIEP) that is defined as the ratio of index-case 
exportation events from the domiciles of a source country to the total number of 
such events ($71$). It may be noted that China has the largest GIEP followed by 
Italy and USA.}
    \label{fig:Rgiep}
\end{figure}

A total of $131 (\approx 65 \%)$  out of $202$ index-cases were found to be 
local nationals. Here we mentioned $202$ index-cases because some countries, 
like, Kuwait, Aruba, Kazakhstan had reported more than one cases on the same 
day (Supplementary Table 1). Remaining $71$ ($45 \%$) index-cases were foreign 
nationals belonging to $22$ countries. Among these foreign nationals, $19$ are 
found to be Chinese followed by $14$ Italians (see Table \ref{tab:ISP}). To 
estimate the global infection exportation potential (GIEP) per country, we 
further normalized the number foreign nationals belongiing to each of the source 
country by the total number of exportation events ($71$) carried out by foreign 
nationals (Fig. \ref{fig:Rgiep}). Above data clearly suggests that although 
most of the infections are imported to the target countries by their domiciles, 
however, a significant proportion of countries are infected by foreign 
nationals. As an example, the number of virus transmission events ($47$) that 
took place from Italy is very high, however, there were only $14$ Italians 
involved in these transmission events.  Thus, in the light of above data, it 
can be inferred that highly robust and efficient transportation system has 
played a huge role in spreading the infection across international borders. 
Therefore, more caution is required specially from aviation departments 
regarding individual's health at international transportation junctions.

\subsection{Connectivity dynamics modeling and analysis of C19-TraNet}

\begin{figure}
     \centering
        
     \begin{subfigure}[b]{0.55\textwidth}
         \includegraphics[width=\textwidth]{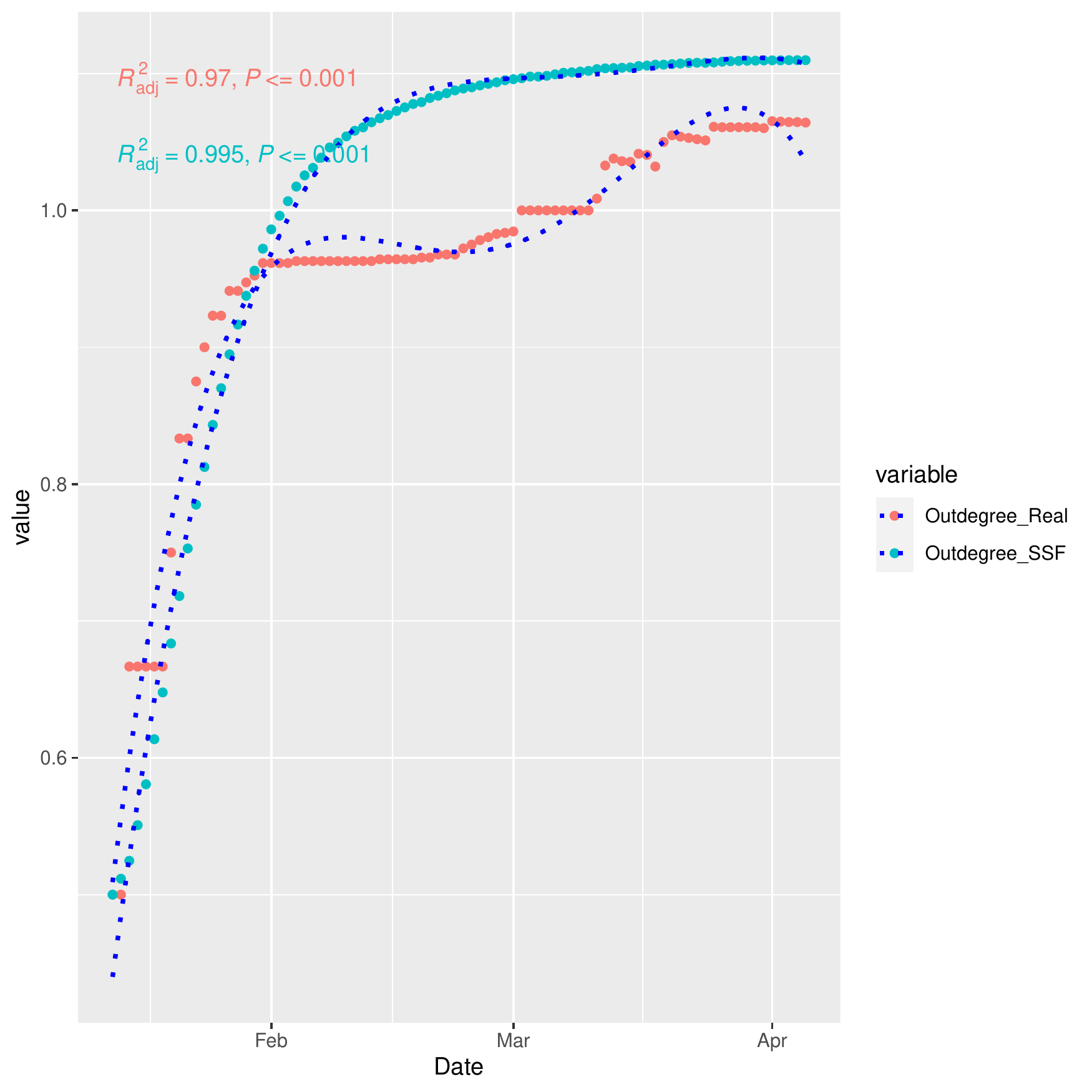}
         \caption{}
         \label{fig:Deg}
     \end{subfigure}
     ~
     \begin{subfigure}[b]{0.55\textwidth}
         \includegraphics[width=\textwidth]{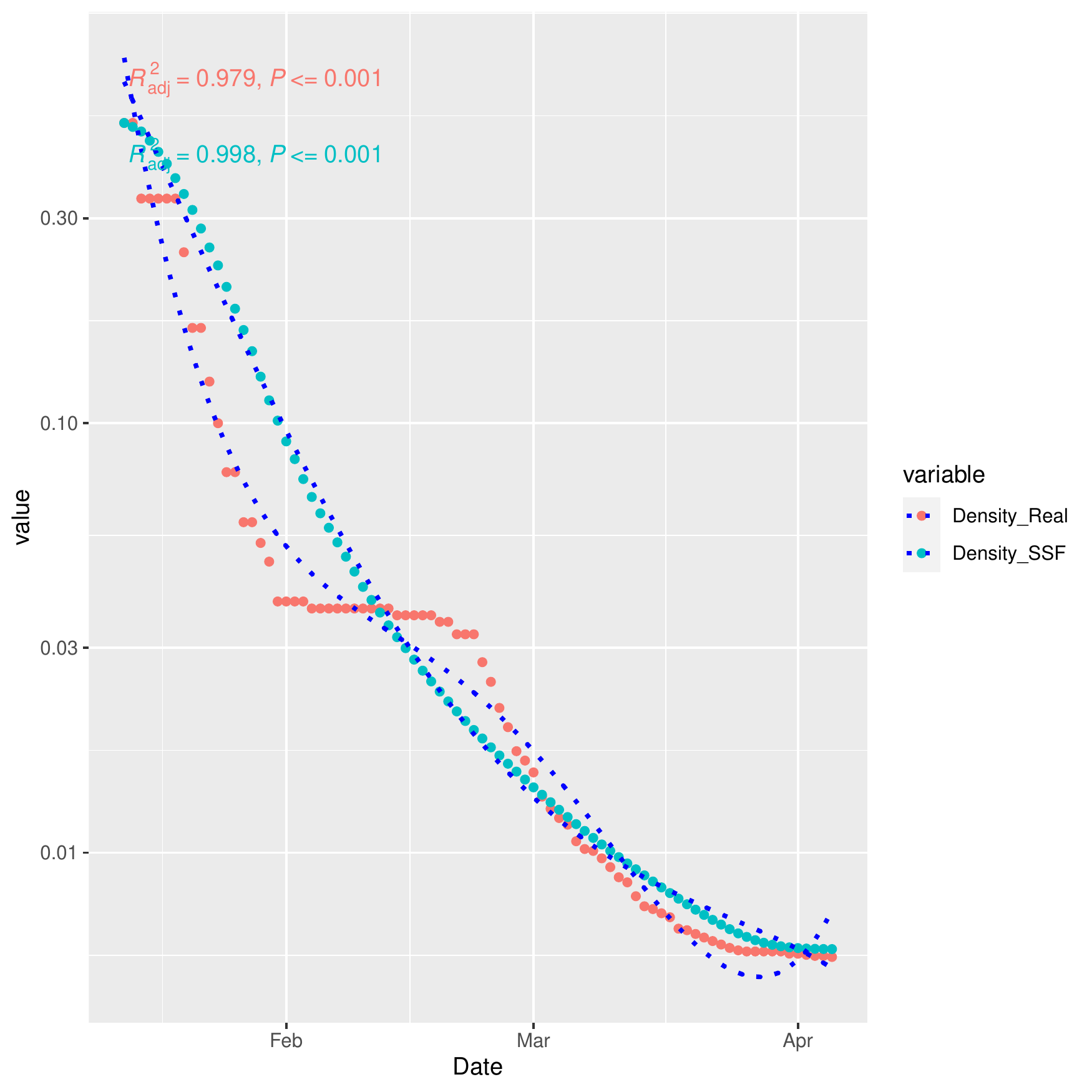}
         \caption{}
         \label{fig:Den}
     \end{subfigure}

     \caption{Disease dynamics of SARS-CoV-2. (a) Comparison of daily 
reported average outdegree (red circles) of growing C19-TraNet with the average of 
outdegree (cyan circles) obtained from an ensemble of $1,000$ stochastic scale 
free realizations, modeled using parameters $\alpha = 0.05$,  $\beta 
= 0.95, \id{maximum-nodes} = 6$ and  $\id{maximum-nodes} = 5$. Blue colored 
dotted line indicates $4^{th}$ order polynomial curve fit.  (b) Comparison of 
daily reported average density of C19-TraNet with average density obtained 
from \hl{$1,000$ SSF models}.}
\end{figure}

Since the pandemic evolves with time, so exploring the connectivity of growing 
C19-TraNet at different instances of time ($t$) can provide critical insights 
about the network topology and growth mechanisms. Keeping that in mind we 
computed average out-degree of C19-TraNet for each of the $84$ days. Further, 
$1,000$ random networks using SSF model corresponding to C19-TraNet were 
generated and average out-degree vectors for all these random networks were 
computed which were further averaged to yield a single vector. We then 
performed Mann-Whitney U test to verify the null hypothesis ($H_0$) that 
the two outdegree distributions (of real network and average of $1,000$ random 
networks) are equal,  which is refuted by a very low $p-value = 3.041e-09$. 
Distributions of average outdegree per unit time (days) for both real and 
average of SSF models are shown in Fig \ref{fig:Deg}. We fitted four 
distributions \textit{viz.} linear, logarithmic, exponential and, polynomial 
(up to order four), among which $4^{th}$ order polynomial curve is found to 
have the highest adjusted coefficient of determination (Adj. $R^2$) for both 
the real and SSF distributions (Table \ref{tab:Trend}). 

In the initial growth phase (\textit{i.e.} till the end of January, 2020), both 
the curves followed same trend, however, from February onwards a peculiar phase 
of no growth (for about a fortnight) is observed  in C19-TraNet which is absent 
in the random networks (Fig \ref{fig:Deg}). This delay emerges due to 
implementation of Wuhan ban and can be attributed to the border control 
measures including local and international travel ban, quarantine, symptom 
screening at airports etc. simultaneously adopted by China and other countries. 
As stated earlier, most of the infections in Chinese wave are disseminated 
from China, therefore the two consecutive travel bans implemented by China 
substantially reduced the rapid local and international flow of virus 
\cite{kraemer2020effect,wells2020impact}. The second surge in average 
connectivity (starting from the end of February, 2020) marks the onset of 
European wave which can be attributed to passage of undocumented pre- or 
asymptomatic infected people from countries other than China where restrictions 
were not that much strict \cite{wells2020impact,li2020substantial}. 
Though the contagion invaded most of the world, however, our model suggests 
that these containment measures have delayed the global transmission of virus 
by $51$ days. If the growth had continued to follow the initial trends, the 
virus would have invaded South Sudan ($187^{th}$ country to be infected) on 
\mydate{2020}{2}{13}. We obtained similar trend, in reverse order, 
characterized by a phase of no growth for approximately same 
period when density is plotted against time (see Fig \ref{fig:Den}). Network 
density peaked when the network was very small. With the passage of 
time, addition of new countries brought small increase in network size in 
comparison to the network order resulting in an overall decrease in network 
density. 

\subsection{Community structure of C19-TraNet} 

\begin{figure}
     \centering
        
     \begin{subfigure}[b]{0.70\textwidth}
         \includegraphics[width=\textwidth]{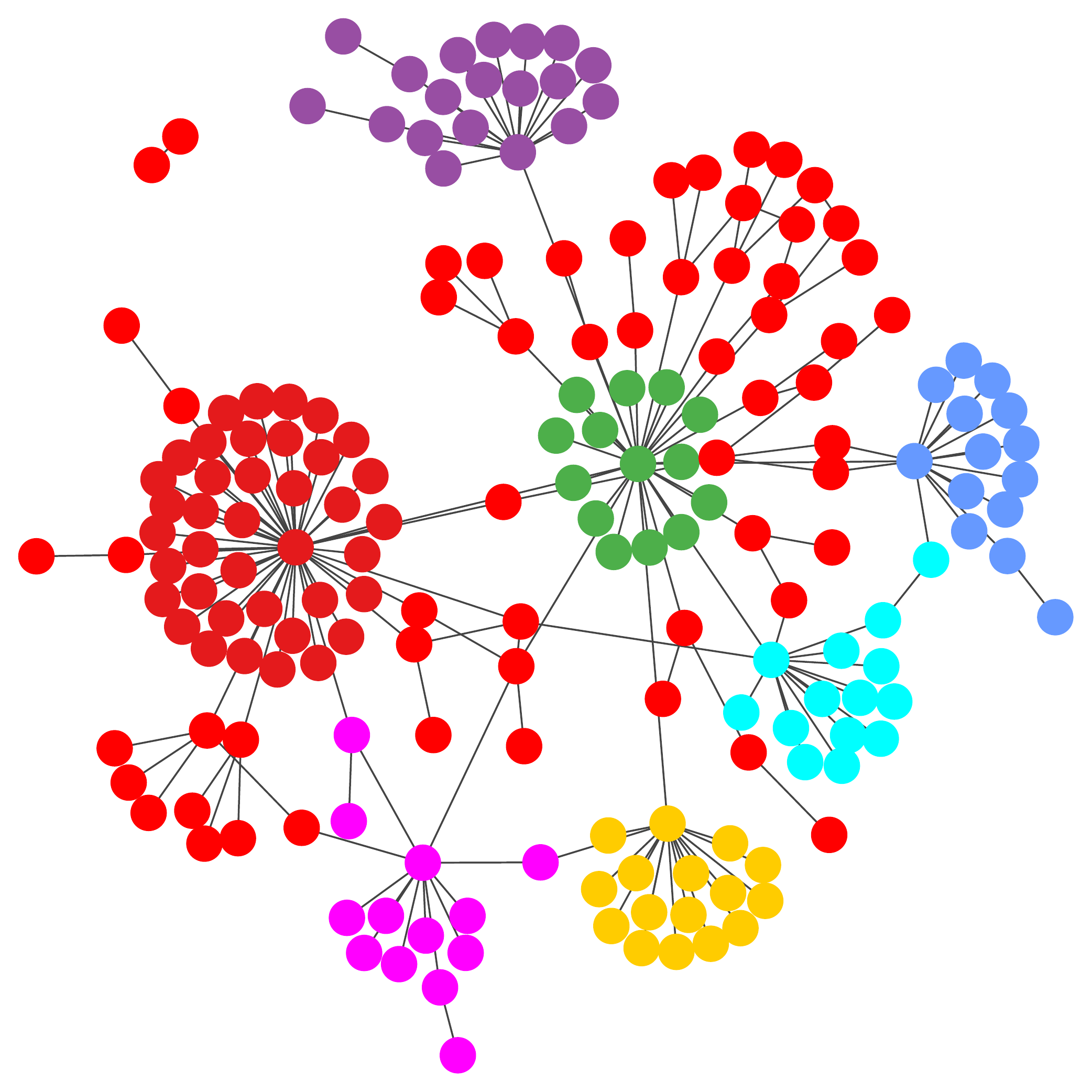}
         \caption{}
         \label{fig:CNet}
     \end{subfigure}
    
     \begin{subfigure}[b]{0.65\textwidth}
         \includegraphics[width=\textwidth]{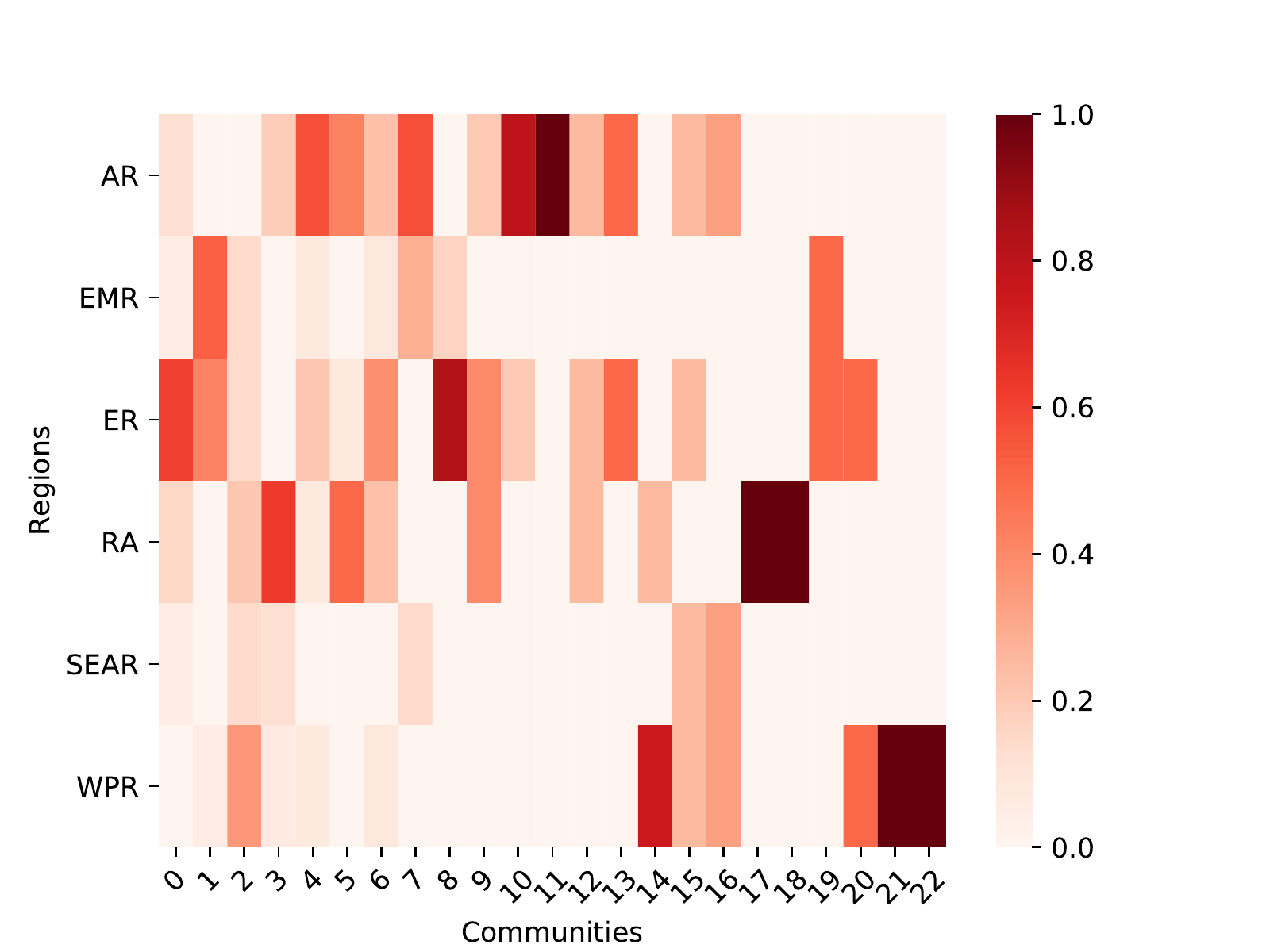}
         \caption{}
         \label{fig:CHmap}
     \end{subfigure}

     \caption{(a) Top seven communities of C19-TraNet based on edge 
betweenness. (b) Percentage of countries shared by individual region per 
community.} 
\end{figure}

We explored the structure of C19-TraNet by leveraging edge betweenness, in 
which  high betweenness nodes were recursively removed to identify sparsely 
connected `node sets' (communities) whose members are densely connected among 
themselves \citep{newman2004finding}. Communities provide an insight about the 
local connectivity patterns in C19-TraNet by partitioning the network into 
densely connected, small sub-graphs. A total of $23$ different communities 
(Fig. \ref{fig:CNet}) were obtained with the largest set consisting of $39$ 
countries followed by $16$ countries  in the second community and $14$ in the 
third set (Table. \ref{ Tab:ComCount}). Most of the communities were small 
sized, while very few consisted of large number of countries (hubs), indicating 
that few countries (having high volume of international traffic) have acted as 
super spreaders. We further grouped the countries into six geographical regions 
as per WHO, namely, European Region (ER), Western Pacific Region (WPR), 
South-East Asia Region (SEAR), Eastern Mediterranean Region (EMR), Region of 
the Americas (RA), African Region (AR) and enumerated the country share of each 
region per community. 

As shown in Fig. \ref{fig:CHmap}, most of communities are composed of a 
heterogeneous mix of countries from different regions. Most of the communities 
are comprised of countries from more than one region among which larger regions 
(in terms of number of countries) including ER, RA and AR are the major 
contributors. Approximately $62\%$ of the countries in Community$\_0$ belongs 
to ER followed by a relatively few countries of RA ($15\%$) and AR ($\approx 
13\%$) regions. Similarly, Community$\_1$ is also comprised mostly of countries 
from two regions, ER ($42\%$) and EM ($\approx 53\%$) (Fig. \ref{fig:CHmap}). 
These results indicate that there is not any kind of positional constraint 
imposed on the spread of infection, \textit{i.e.}, infected countries from a 
given region have spread the infection to their neighboring countries (short 
range interactions) as well as across non neighboring countries or territories 
which are geographically far away from them (long range interactions).

\section{Summary and Conclusion}

In this  study, we have constructed the an empirical index-case global 
transmission network of infection caused by SARS-CoV-2. Our main objective here 
is to develop a systems level understanding of emergence and growth of COVID-19 
from a local 
outbreak into a global pandemic.
To achieve this goal, we relied on the travel history of index-case patients 
obtained from 
various reliable sources. It allowed us to map multiple transmission routes 
through 
which the virus intruded into different geographical locations. 

We observe that, after its origin in China, the global transmission of 
SARS-CoV-2 can be attributed to two waves, namely, the Chinese wave and the 
European wave. Initially, during Chinese wave, the virus invaded to a large 
number of neighboring as well as distantly 
located countries from China due to a large volume of international traffic. By 
the end of January, 2020, the effect of Chinese wave started to diminish due to 
implementation of various border control measures within and outside China. 
However, soon after a small stagnation phase, the virus started to spread from 
other infected countries 
mainly from European Regions. The main reason of this European wave was that a 
large number of asymptomatic patients (local or foreign nationals) passed 
undetected to different countries and by the time we responded, it had 
already victimized the whole world. 

Topological structure of C19-TraNet is found to be sparse, however, its degree 
distribution follows a power law 
distribution, \textit{i.e.}, it possess scale free property suggesting that few 
countries 
have played the role of super spreaders in virus transmission. Considering the 
C19-TraNet as an evolving network, we further modeled the temporal growth 
of its average outdegree and found a delay of $51$ days in virus transmission 
that can be attributed to different border control measures. Data suggests that 
large incubation period, free flow of international traffic and lack of 
coordinated action at international level in implementing various control 
strategies have played a vital role in global transmission of the virus. 
Exploration of community structure of the 
network revealed that large communities are not comprised of countries from a 
single region, however, they are heterogeneous composition of countries from 
different 
regions suggesting the presence of local transmission, to neighboring countries 
of the same region, as well as long range transmission to countries belonging 
to different regions. These multiple transmission paths allowed the virus to 
rapidly invade susceptible subpopulations located in distant geographical 
regions. This conclusion emphasizes the imposition of travel restrictions which 
although implemented, however, lack coordinated actions from different 
countries. It suggests that a coordinated action plan is needed to put in 
places to fight with any future outbreak.

The stochastic scale free (SSF) algorithm, proposed in this work to model the 
first 
case global transmission network of SARS-CoV-2, can be implemented very easily 
to model other  human-to-human transmission 
disease networks with minor changes. The main advantage of the proposed 
algorithm is that the 
probability of infecting new nodes changes with change in number of infected 
nodes, that increases continuously until the network order 
approaches saturation.

\begin{acknowledgement}
We thank Central University of Himachal Pradesh for providing the required 
infrastructure and computational facilities. $\text{VS}^{\dagger}$ thanks 
Council of Scientific and Industrial Research (CSIR), India for providing 
Junior Research Fellowship (JRF). 
\end{acknowledgement}

\section{Author Contributions}
$\text{VS}^*$ conceptualized and designed the research framework. 
$\text{VS}^{\dagger}$ performed the computational 
experiments. $\text{VS}^{\dagger}$ and $\text{VS}^*$ analyzed the data and 
interpreted results. $\text{VS}^{\dagger}$ and $\text{VS}^*$ wrote and 
finalized the manuscript.

\section{Conflict of Interests}
The authors declare that they have no conflict of interests.

\bibliography{CoV_TrNet}

\providecommand{\latin}[1]{#1}
\makeatletter
\providecommand{\doi}
  {\begingroup\let\do\@makeother\dospecials
  \catcode`\{=1 \catcode`\}=2 \doi@aux}
\providecommand{\doi@aux}[1]{\endgroup\texttt{#1}}
\makeatother
\providecommand*\mcitethebibliography{\thebibliography}
\csname @ifundefined\endcsname{endmcitethebibliography}
  {\let\endmcitethebibliography\endthebibliography}{}
\begin{mcitethebibliography}{44}
\providecommand*\natexlab[1]{#1}
\providecommand*\mciteSetBstSublistMode[1]{}
\providecommand*\mciteSetBstMaxWidthForm[2]{}
\providecommand*\mciteBstWouldAddEndPuncttrue
  {\def\EndOfBibitem{\unskip.}}
\providecommand*\mciteBstWouldAddEndPunctfalse
  {\let\EndOfBibitem\relax}
\providecommand*\mciteSetBstMidEndSepPunct[3]{}
\providecommand*\mciteSetBstSublistLabelBeginEnd[3]{}
\providecommand*\EndOfBibitem{}
\mciteSetBstSublistMode{f}
\mciteSetBstMaxWidthForm{subitem}{(\alph{mcitesubitemcount})}
\mciteSetBstSublistLabelBeginEnd
  {\mcitemaxwidthsubitemform\space}
  {\relax}
  {\relax}

\bibitem[Gorbalenya \latin{et~al.}(2020)Gorbalenya, Baker, Baric, de~Groot,
  Drosten, Gulyaeva, Haagmans, Lauber, Leontovich, Neuman, Penzar, Perlman,
  Poon, Samborskiy, Sidorov, Sola, and Ziebuhr]{of2020species}
Gorbalenya,~A.~E. \latin{et~al.}  The species Severe acute respiratory
  syndrome-related coronavirus: classifying 2019-nCoV and naming it SARS-CoV-2.
  \emph{Nature Microbiology} \textbf{2020}, \emph{5}, 536--544\relax
\mciteBstWouldAddEndPuncttrue
\mciteSetBstMidEndSepPunct{\mcitedefaultmidpunct}
{\mcitedefaultendpunct}{\mcitedefaultseppunct}\relax
\EndOfBibitem
\bibitem[Zhu \latin{et~al.}(2020)Zhu, Zhang, Wang, Li, Yang, Song, Zhao, Huang,
  Shi, Lu, Niu, Zhan, Ma, Wang, Xu, Wu, Gao, and Tan]{zhu2020novel}
Zhu,~N. \latin{et~al.}  A novel coronavirus from patients with pneumonia in
  China, 2019. \emph{New England Journal of Medicine} \textbf{2020},
  \emph{382}, 727--733\relax
\mciteBstWouldAddEndPuncttrue
\mciteSetBstMidEndSepPunct{\mcitedefaultmidpunct}
{\mcitedefaultendpunct}{\mcitedefaultseppunct}\relax
\EndOfBibitem
\bibitem[Organization(2020)]{world2005statement}
Organization,~W.~H. Statement on the second meeting of the International Health
  Regulations (2005) Emergency Committee regarding the outbreak of novel
  coronavirus (2019-nCoV). 2020;
  \url{https://www.who.int/news-room/detail/30-01-2020-statement-on-the-second-meeting-of-the-international-health-regulations-(2005)-emergency-committee-regarding-the-outbreak-of-novel-coronavirus-(2019-ncov)}\relax
\mciteBstWouldAddEndPuncttrue
\mciteSetBstMidEndSepPunct{\mcitedefaultmidpunct}
{\mcitedefaultendpunct}{\mcitedefaultseppunct}\relax
\EndOfBibitem
\bibitem[Organization(2020)]{world2020director}
Organization,~W.~H. WHO Director-General's opening remarks at the media
  briefing on COVID-19-11 March 2020. 2020;
  \url{https://www.who.int/dg/speeches/detail/who-director-general-s-opening-remarks-at-the-media-briefing-on-covid-19---11-march-2020}\relax
\mciteBstWouldAddEndPuncttrue
\mciteSetBstMidEndSepPunct{\mcitedefaultmidpunct}
{\mcitedefaultendpunct}{\mcitedefaultseppunct}\relax
\EndOfBibitem
\bibitem[Organization(2020)]{WHOSR109}
Organization,~W.~H. Coronavirus disease (COVID-19): situation report-109. 2020;
  \url{https://www.who.int/docs/default-source/coronaviruse/situation-reports/20200508covid-19-sitrep-109.pdf?sfvrsn=68f2c632_6}\relax
\mciteBstWouldAddEndPuncttrue
\mciteSetBstMidEndSepPunct{\mcitedefaultmidpunct}
{\mcitedefaultendpunct}{\mcitedefaultseppunct}\relax
\EndOfBibitem
\bibitem[Wang \latin{et~al.}(2020)Wang, Hu, Hu, Zhu, Liu, Zhang, Wang, Xiang,
  Cheng, Xiong, \latin{et~al.} others]{wang2020clinical}
Wang,~D. \latin{et~al.}  Clinical characteristics of 138 hospitalized patients
  with 2019 novel coronavirus--infected pneumonia in Wuhan, China. \emph{Jama}
  \textbf{2020}, \emph{323}, 1061--1069\relax
\mciteBstWouldAddEndPuncttrue
\mciteSetBstMidEndSepPunct{\mcitedefaultmidpunct}
{\mcitedefaultendpunct}{\mcitedefaultseppunct}\relax
\EndOfBibitem
\bibitem[Bai \latin{et~al.}(2020)Bai, Yao, Wei, Tian, Jin, Chen, and
  Wang]{bai2020presumed}
Bai,~Y. \latin{et~al.}  Presumed asymptomatic carrier transmission of COVID-19.
  \emph{Jama} \textbf{2020}, \emph{323}, 1406--1407\relax
\mciteBstWouldAddEndPuncttrue
\mciteSetBstMidEndSepPunct{\mcitedefaultmidpunct}
{\mcitedefaultendpunct}{\mcitedefaultseppunct}\relax
\EndOfBibitem
\bibitem[Tatem \latin{et~al.}(2006)Tatem, Rogers, and Hay]{tatem2006global}
Tatem,~A.~J.; Rogers,~D.~J.; Hay,~S.~I. Global transport networks and
  infectious disease spread. \emph{Advances in parasitology} \textbf{2006},
  \emph{62}, 293--343\relax
\mciteBstWouldAddEndPuncttrue
\mciteSetBstMidEndSepPunct{\mcitedefaultmidpunct}
{\mcitedefaultendpunct}{\mcitedefaultseppunct}\relax
\EndOfBibitem
\bibitem[Fidler(1996)]{fidler1996globalization}
Fidler,~D.~P. Globalization, international law, and emerging infectious
  diseases. \emph{Emerging infectious diseases} \textbf{1996}, \emph{2},
  77\relax
\mciteBstWouldAddEndPuncttrue
\mciteSetBstMidEndSepPunct{\mcitedefaultmidpunct}
{\mcitedefaultendpunct}{\mcitedefaultseppunct}\relax
\EndOfBibitem
\bibitem[Keeling and Gilligan(2000)Keeling, and
  Gilligan]{keeling2000metapopulation}
Keeling,~M.~J.; Gilligan,~C.~A. Metapopulation dynamics of bubonic plague.
  \emph{Nature} \textbf{2000}, \emph{407}, 903--906\relax
\mciteBstWouldAddEndPuncttrue
\mciteSetBstMidEndSepPunct{\mcitedefaultmidpunct}
{\mcitedefaultendpunct}{\mcitedefaultseppunct}\relax
\EndOfBibitem
\bibitem[Sack(2004)]{sack2004sack}
Sack,~D.~A. Sack RB, Nair GB, and Siddique AK. \emph{Cholera. Lancet}
  \textbf{2004}, \emph{363}, 223--233\relax
\mciteBstWouldAddEndPuncttrue
\mciteSetBstMidEndSepPunct{\mcitedefaultmidpunct}
{\mcitedefaultendpunct}{\mcitedefaultseppunct}\relax
\EndOfBibitem
\bibitem[Ferguson \latin{et~al.}(2003)Ferguson, Galvani, and
  Bush]{ferguson2003ecological}
Ferguson,~N.~M.; Galvani,~A.~P.; Bush,~R.~M. Ecological and immunological
  determinants of influenza evolution. \emph{Nature} \textbf{2003}, \emph{422},
  428--433\relax
\mciteBstWouldAddEndPuncttrue
\mciteSetBstMidEndSepPunct{\mcitedefaultmidpunct}
{\mcitedefaultendpunct}{\mcitedefaultseppunct}\relax
\EndOfBibitem
\bibitem[Perrin \latin{et~al.}(2003)Perrin, Kaiser, and
  Yerly]{perrin2003travel}
Perrin,~L.; Kaiser,~L.; Yerly,~S. Travel and the spread of HIV-1 genetic
  variants. \emph{The Lancet infectious diseases} \textbf{2003}, \emph{3},
  22--27\relax
\mciteBstWouldAddEndPuncttrue
\mciteSetBstMidEndSepPunct{\mcitedefaultmidpunct}
{\mcitedefaultendpunct}{\mcitedefaultseppunct}\relax
\EndOfBibitem
\bibitem[Peiris \latin{et~al.}(2004)Peiris, Guan, and Yuen]{peiris2004severe}
Peiris,~J.; Guan,~Y.; Yuen,~K. Severe acute respiratory syndrome. \emph{Nature
  medicine} \textbf{2004}, \emph{10}, S88--S97\relax
\mciteBstWouldAddEndPuncttrue
\mciteSetBstMidEndSepPunct{\mcitedefaultmidpunct}
{\mcitedefaultendpunct}{\mcitedefaultseppunct}\relax
\EndOfBibitem
\bibitem[Zinsser and Grob(2011)Zinsser, and Grob]{zinsser1935rats}
Zinsser,~H.; Grob,~G. \emph{Rats, Lice and History}; Social Science Classics;
  Transaction Publishers: Piscataway, New Jersey, United States, 2011\relax
\mciteBstWouldAddEndPuncttrue
\mciteSetBstMidEndSepPunct{\mcitedefaultmidpunct}
{\mcitedefaultendpunct}{\mcitedefaultseppunct}\relax
\EndOfBibitem
\bibitem[Reidl and Klose(2002)Reidl, and Klose]{reidl2002vibrio}
Reidl,~J.; Klose,~K.~E. Vibrio cholerae and cholera: out of the water and into
  the host. \emph{FEMS microbiology reviews} \textbf{2002}, \emph{26},
  125--139\relax
\mciteBstWouldAddEndPuncttrue
\mciteSetBstMidEndSepPunct{\mcitedefaultmidpunct}
{\mcitedefaultendpunct}{\mcitedefaultseppunct}\relax
\EndOfBibitem
\bibitem[Palese(2004)]{palese2004influenza}
Palese,~P. Influenza: old and new threats. \emph{Nature medicine}
  \textbf{2004}, \emph{10}, S82--S87\relax
\mciteBstWouldAddEndPuncttrue
\mciteSetBstMidEndSepPunct{\mcitedefaultmidpunct}
{\mcitedefaultendpunct}{\mcitedefaultseppunct}\relax
\EndOfBibitem
\bibitem[Heymann(2004)]{heymann2004international}
Heymann,~D.~L. The international response to the outbreak of SARS in 2003.
  \emph{Philosophical Transactions of the Royal Society of London. Series B:
  Biological Sciences} \textbf{2004}, \emph{359}, 1127--1129\relax
\mciteBstWouldAddEndPuncttrue
\mciteSetBstMidEndSepPunct{\mcitedefaultmidpunct}
{\mcitedefaultendpunct}{\mcitedefaultseppunct}\relax
\EndOfBibitem
\bibitem[Bloom and Cadarette(2019)Bloom, and Cadarette]{bloom2019infectious}
Bloom,~D.~E.; Cadarette,~D. Infectious Disease Threats in the 21st Century:
  Strengthening the Global Response. \emph{Frontiers in immunology}
  \textbf{2019}, \emph{10}, 549\relax
\mciteBstWouldAddEndPuncttrue
\mciteSetBstMidEndSepPunct{\mcitedefaultmidpunct}
{\mcitedefaultendpunct}{\mcitedefaultseppunct}\relax
\EndOfBibitem
\bibitem[Epstein(2009)]{epstein2009modelling}
Epstein,~J.~M. Modelling to contain pandemics. \emph{Nature} \textbf{2009},
  \emph{460}, 687--687\relax
\mciteBstWouldAddEndPuncttrue
\mciteSetBstMidEndSepPunct{\mcitedefaultmidpunct}
{\mcitedefaultendpunct}{\mcitedefaultseppunct}\relax
\EndOfBibitem
\bibitem[Gates(2018)]{gates2018innovation}
Gates,~B. Innovation for pandemics. \emph{New England Journal of Medicine}
  \textbf{2018}, \emph{378}, 2057--2060\relax
\mciteBstWouldAddEndPuncttrue
\mciteSetBstMidEndSepPunct{\mcitedefaultmidpunct}
{\mcitedefaultendpunct}{\mcitedefaultseppunct}\relax
\EndOfBibitem
\bibitem[Weiss(2003)]{weiss2003hiv}
Weiss,~R.~A. HIV and AIDS in relation to other pandemics. \emph{EMBO reports}
  \textbf{2003}, \emph{4}, S10--S14\relax
\mciteBstWouldAddEndPuncttrue
\mciteSetBstMidEndSepPunct{\mcitedefaultmidpunct}
{\mcitedefaultendpunct}{\mcitedefaultseppunct}\relax
\EndOfBibitem
\bibitem[Salath{\'e} \latin{et~al.}(2010)Salath{\'e}, Kazandjieva, Lee, Levis,
  Feldman, and Jones]{salathe2010high}
Salath{\'e},~M. \latin{et~al.}  A high-resolution human contact network for
  infectious disease transmission. \emph{Proceedings of the National Academy of
  Sciences} \textbf{2010}, \emph{107}, 22020--22025\relax
\mciteBstWouldAddEndPuncttrue
\mciteSetBstMidEndSepPunct{\mcitedefaultmidpunct}
{\mcitedefaultendpunct}{\mcitedefaultseppunct}\relax
\EndOfBibitem
\bibitem[Enright and Kao(2018)Enright, and Kao]{enright2018epidemics}
Enright,~J.; Kao,~R.~R. Epidemics on dynamic networks. \emph{Epidemics}
  \textbf{2018}, \emph{24}, 88--97\relax
\mciteBstWouldAddEndPuncttrue
\mciteSetBstMidEndSepPunct{\mcitedefaultmidpunct}
{\mcitedefaultendpunct}{\mcitedefaultseppunct}\relax
\EndOfBibitem
\bibitem[Keeling and Eames(2005)Keeling, and Eames]{keeling2005networks}
Keeling,~M.~J.; Eames,~K.~T. Networks and epidemic models. \emph{Journal of the
  Royal Society Interface} \textbf{2005}, \emph{2}, 295--307\relax
\mciteBstWouldAddEndPuncttrue
\mciteSetBstMidEndSepPunct{\mcitedefaultmidpunct}
{\mcitedefaultendpunct}{\mcitedefaultseppunct}\relax
\EndOfBibitem
\bibitem[Craft(2015)]{craft2015infectious}
Craft,~M.~E. Infectious disease transmission and contact networks in wildlife
  and livestock. \emph{Philosophical Transactions of the Royal Society B:
  Biological Sciences} \textbf{2015}, \emph{370}, 20140107\relax
\mciteBstWouldAddEndPuncttrue
\mciteSetBstMidEndSepPunct{\mcitedefaultmidpunct}
{\mcitedefaultendpunct}{\mcitedefaultseppunct}\relax
\EndOfBibitem
\bibitem[Bell \latin{et~al.}(1999)Bell, Atkinson, and
  Carlson]{bell1999centrality}
Bell,~D.~C.; Atkinson,~J.~S.; Carlson,~J.~W. Centrality measures for disease
  transmission networks. \emph{Social networks} \textbf{1999}, \emph{21},
  1--21\relax
\mciteBstWouldAddEndPuncttrue
\mciteSetBstMidEndSepPunct{\mcitedefaultmidpunct}
{\mcitedefaultendpunct}{\mcitedefaultseppunct}\relax
\EndOfBibitem
\bibitem[Newman(2003)]{Newman2003}
Newman,~M. E.~J. {The structure and function of complex networks}.
  \textbf{2003}, \emph{45}, 167--256\relax
\mciteBstWouldAddEndPuncttrue
\mciteSetBstMidEndSepPunct{\mcitedefaultmidpunct}
{\mcitedefaultendpunct}{\mcitedefaultseppunct}\relax
\EndOfBibitem
\bibitem[Watts and Strogatz(1998)Watts, and Strogatz]{watts1998collective}
Watts,~D.~J.; Strogatz,~S.~H. Collective dynamics of small-world networks.
  \emph{Nature} \textbf{1998}, \emph{393}, 440\relax
\mciteBstWouldAddEndPuncttrue
\mciteSetBstMidEndSepPunct{\mcitedefaultmidpunct}
{\mcitedefaultendpunct}{\mcitedefaultseppunct}\relax
\EndOfBibitem
\bibitem[Newman and Girvan(2004)Newman, and Girvan]{newman2004finding}
Newman,~M.~E.; Girvan,~M. Finding and evaluating community structure in
  networks. \emph{Physical review E} \textbf{2004}, \emph{69}, 026113\relax
\mciteBstWouldAddEndPuncttrue
\mciteSetBstMidEndSepPunct{\mcitedefaultmidpunct}
{\mcitedefaultendpunct}{\mcitedefaultseppunct}\relax
\EndOfBibitem
\bibitem[Mann and Whitney(1947)Mann, and Whitney]{mann1947test}
Mann,~H.~B.; Whitney,~D.~R. On a test of whether one of two random variables is
  stochastically larger than the other. \emph{The annals of mathematical
  statistics} \textbf{1947}, 50--60\relax
\mciteBstWouldAddEndPuncttrue
\mciteSetBstMidEndSepPunct{\mcitedefaultmidpunct}
{\mcitedefaultendpunct}{\mcitedefaultseppunct}\relax
\EndOfBibitem
\bibitem[Mangili and Gendreau(2005)Mangili, and Gendreau]{Mangili2005}
Mangili,~A.; Gendreau,~M.~A. Transmission of infectious diseases during
  commercial air travel. \emph{The Lancet} \textbf{2005}, \emph{365},
  0140--6736\relax
\mciteBstWouldAddEndPuncttrue
\mciteSetBstMidEndSepPunct{\mcitedefaultmidpunct}
{\mcitedefaultendpunct}{\mcitedefaultseppunct}\relax
\EndOfBibitem
\bibitem[Trindade~d’{\'A}vila Magalh{\~a}es(2018)]{trindade2018globaliser}
Trindade~d’{\'A}vila Magalh{\~a}es,~D. The globaliser dragon: how is China
  changing economic globalisation? \emph{Third World Quarterly} \textbf{2018},
  \emph{39}, 1727--1749\relax
\mciteBstWouldAddEndPuncttrue
\mciteSetBstMidEndSepPunct{\mcitedefaultmidpunct}
{\mcitedefaultendpunct}{\mcitedefaultseppunct}\relax
\EndOfBibitem
\bibitem[Daszak \latin{et~al.}(2020)Daszak, Olival, and Li]{daszak2020strategy}
Daszak,~P.; Olival,~K.~J.; Li,~H. A strategy to prevent future pandemics
  similar to the 2019-nCoV outbreak. \emph{Biosafety and Health} \textbf{2020},
  \emph{2}, 6--8\relax
\mciteBstWouldAddEndPuncttrue
\mciteSetBstMidEndSepPunct{\mcitedefaultmidpunct}
{\mcitedefaultendpunct}{\mcitedefaultseppunct}\relax
\EndOfBibitem
\bibitem[Zhao \latin{et~al.}(2020)Zhao, Zhuang, Ran, Lin, Yang, Yang, and
  He]{zhao2020association}
Zhao,~S. \latin{et~al.}  The association between domestic train transportation
  and novel coronavirus (2019-nCoV) outbreak in China from 2019 to 2020: a
  data-driven correlational report. \emph{Travel medicine and infectious
  disease} \textbf{2020}, \emph{33}, 101568\relax
\mciteBstWouldAddEndPuncttrue
\mciteSetBstMidEndSepPunct{\mcitedefaultmidpunct}
{\mcitedefaultendpunct}{\mcitedefaultseppunct}\relax
\EndOfBibitem
\bibitem[Bogoch \latin{et~al.}(2020)Bogoch, Watts, Thomas-Bachli, Huber,
  Kraemer, and Khan]{bogoch2020pneumonia}
Bogoch,~I.~I. \latin{et~al.}  Pneumonia of Unknown Etiology in Wuhan, China:
  Potential for International Spread Via Commercial Air Travel. \emph{Journal
  of Travel Medicine} \textbf{2020}, \emph{27}, taaa008\relax
\mciteBstWouldAddEndPuncttrue
\mciteSetBstMidEndSepPunct{\mcitedefaultmidpunct}
{\mcitedefaultendpunct}{\mcitedefaultseppunct}\relax
\EndOfBibitem
\bibitem[Organization(2020)]{WHOSR5}
Organization,~W.~H. Novel Coronavirus (2019-nCoV): situation report-5. 2020;
  \url{https://www.who.int/docs/default-source/coronaviruse/situation-reports/20200125-sitrep-5-2019-ncov.pdf?sfvrsn=429b143d_8}\relax
\mciteBstWouldAddEndPuncttrue
\mciteSetBstMidEndSepPunct{\mcitedefaultmidpunct}
{\mcitedefaultendpunct}{\mcitedefaultseppunct}\relax
\EndOfBibitem
\bibitem[Chinazzi \latin{et~al.}(2020)Chinazzi, Davis, Ajelli, Gioannini,
  Litvinova, Merler, y~Piontti, Mu, Rossi, Sun, \latin{et~al.}
  others]{chinazzi2020effect}
Chinazzi,~M. \latin{et~al.}  The effect of travel restrictions on the spread of
  the 2019 novel coronavirus (COVID-19) outbreak. \emph{Science} \textbf{2020},
  \emph{368}, 395--400\relax
\mciteBstWouldAddEndPuncttrue
\mciteSetBstMidEndSepPunct{\mcitedefaultmidpunct}
{\mcitedefaultendpunct}{\mcitedefaultseppunct}\relax
\EndOfBibitem
\bibitem[Li \latin{et~al.}(2020)Li, Pei, Chen, Song, Zhang, Yang, and
  Shaman]{li2020substantial}
Li,~R. \latin{et~al.}  Substantial undocumented infection facilitates the rapid
  dissemination of novel coronavirus (SARS-CoV-2). \emph{Science}
  \textbf{2020}, \emph{368}, 489--493\relax
\mciteBstWouldAddEndPuncttrue
\mciteSetBstMidEndSepPunct{\mcitedefaultmidpunct}
{\mcitedefaultendpunct}{\mcitedefaultseppunct}\relax
\EndOfBibitem
\bibitem[Rothe \latin{et~al.}(2020)Rothe, Schunk, Sothmann, Bretzel, Froeschl,
  Wallrauch, Zimmer, Thiel, Janke, Guggemos, \latin{et~al.}
  others]{rothe2020transmission}
Rothe,~C. \latin{et~al.}  Transmission of 2019-nCoV infection from an
  asymptomatic contact in Germany. \emph{New England Journal of Medicine}
  \textbf{2020}, \emph{382}, 970--971\relax
\mciteBstWouldAddEndPuncttrue
\mciteSetBstMidEndSepPunct{\mcitedefaultmidpunct}
{\mcitedefaultendpunct}{\mcitedefaultseppunct}\relax
\EndOfBibitem
\bibitem[Schneeberger \latin{et~al.}(2004)Schneeberger, Mercer, Gregson,
  Ferguson, Nyamukapa, Anderson, Johnson, and Garnett]{schneeberger2004scale}
Schneeberger,~A. \latin{et~al.}  Scale-free networks and sexually transmitted
  diseases: a description of observed patterns of sexual contacts in Britain
  and Zimbabwe. \emph{Sexually transmitted diseases} \textbf{2004}, \emph{31},
  380--387\relax
\mciteBstWouldAddEndPuncttrue
\mciteSetBstMidEndSepPunct{\mcitedefaultmidpunct}
{\mcitedefaultendpunct}{\mcitedefaultseppunct}\relax
\EndOfBibitem
\bibitem[Kraemer \latin{et~al.}(2020)Kraemer, Yang, Gutierrez, Wu, Klein,
  Pigott, Du~Plessis, Faria, Li, Hanage, \latin{et~al.}
  others]{kraemer2020effect}
Kraemer,~M.~U. \latin{et~al.}  The effect of human mobility and control
  measures on the COVID-19 epidemic in China. \emph{Science} \textbf{2020},
  \emph{368}, 493--497\relax
\mciteBstWouldAddEndPuncttrue
\mciteSetBstMidEndSepPunct{\mcitedefaultmidpunct}
{\mcitedefaultendpunct}{\mcitedefaultseppunct}\relax
\EndOfBibitem
\bibitem[Wells \latin{et~al.}(2020)Wells, Sah, Moghadas, Pandey, Shoukat, Wang,
  Wang, Meyers, Singer, and Galvani]{wells2020impact}
Wells,~C.~R. \latin{et~al.}  Impact of international travel and border control
  measures on the global spread of the novel 2019 coronavirus outbreak.
  \emph{Proceedings of the National Academy of Sciences} \textbf{2020},
  \emph{117}, 7504--7509\relax
\mciteBstWouldAddEndPuncttrue
\mciteSetBstMidEndSepPunct{\mcitedefaultmidpunct}
{\mcitedefaultendpunct}{\mcitedefaultseppunct}\relax
\EndOfBibitem
\end{mcitethebibliography}

\pagebreak

\begin{table}[htbp]
\caption{Number of countries per region present in individual community.}
\begin{tabular}{r r r r r r r}
\hline
\multicolumn{1}{p{2cm}}{Community} & \multicolumn{1}{p{2cm}}{European Region} & 
\multicolumn{1}{p{2cm}}{Western Pacific Region} & 
\multicolumn{1}{p{2cm}}{South-East Asia Region} & 
\multicolumn{1}{p{2cm}}{Eastern Mediterranean Region} & 
\multicolumn{1}{p{2cm}}{Region of the Americas} & 
\multicolumn{1}{p{2cm}}{African Region} \\ \hline
1 & 24 & 0 & 2 & 2 & 6 & 5 \\
2 & 8 & 1 & 0 & 10 & 0 & 0 \\
3 & 2 & 5 & 2 & 2 & 3 & 0 \\
4 & 0 & 1 & 2 & 0 & 10 & 3 \\
5 & 3 & 1 & 0 & 1 & 1 & 8 \\
6 & 1 & 0 & 0 & 0 & 7 & 6 \\
7 & 5 & 1 & 0 & 1 & 3 & 3 \\ \hline
\end{tabular}
\label{tab:ComCount}
\end{table}

\begin{table}[htbp]
\caption{Network metrics of countries with at least one outdegree.}
\begin{tabular}{l r r r}
\hline
\multicolumn{1}{p{2cm}}{\textbf{Country}} & 
\multicolumn{1}{l}{\textbf{$k_{in}$}} & 
\multicolumn{1}{l}{\textbf{$k_{out}$}} & \multicolumn{1}{l}{\textbf{$L$}} 
\\ \hline
Italy & 1 & 47 & 1.2034 \\
Spain & 1 & 11 & 1.1538 \\
Iran & 1 & 16 & 1.1111 \\
French Republic & 1 & 15 & 1.1176 \\
United States of America & 1 & 16 & 1.0000 \\
United Kingdom & 1 & 14 & 1.0667 \\
Germany & 1 & 4 & 1.8824 \\
Switzerland & 1 & 4 & 1.0000 \\
Netherlands & 1 & 3 & 1.0000 \\
Portugal & 2 & 1 & 1.0000 \\
Japan & 1 & 3 & 1.0000 \\
United Arab Emirates & 1 & 3 & 1.0000 \\
India & 1 & 3 & 1.2500 \\
Belgium & 1 & 3 & 1.2500 \\
Singapore & 1 & 2 & 1.3333 \\
Ecuador & 1 & 1 & 1.0000 \\
Norway & 1 & 2 & 1.3333 \\
Ghana & 2 & 1 & 1.0000 \\
Greece & 1 & 2 & 1.0000 \\
Malaysia & 1 & 1 & 1.0000 \\
Dominican Republic & 1 & 1 & 1.0000 \\
Morocco & 1 & 1 & 1.0000 \\
Saudi Arabia & 1 & 1 & 1.0000 \\
Hungary & 1 & 1 & 1.0000 \\
Cameroon & 1 & 1 & 1.0000 \\
Thailand & 1 & 2 & 1.0000 \\
Côte d’Ivoire & 1 & 1 & 1.0000 \\
Australia & 1 & 1 & 1.0000 \\
Philippines & 1 & 1 & 1.0000 \\
Russian Federation & 1 & 1 & 1.0000 \\
Democratic Republic of the Congo & 1 & 1 & 1.0000 \\
Rwanda & 1 & 1 & 1.0000 \\
Mauritius & 1 & 1 & 1.0000 \\
Togo & 1 & 1 & 1.0000 \\
Burkina Faso & 1 & 1 & 1.0000 \\
China & 0 & 29 & 1.9836 \\
Turkey & 0 & 1 & 1.5000 \\
Panama & 0 & 1 & 1.0000 \\ \hline
\end{tabular}
\label{tab:Prop}
\end{table}

\begin{table}[htbp]
\caption{Number of domiciles of source country spreading infection to target countries and 
global infection exportation potential (GIEP) obtained by normalizing individual number by 
total out degree.}
\begin{tabular}{l r r}
\hline
\multicolumn{1}{p{2cm}}{\textbf{Countries} }& 
\multicolumn{1}{p{2.5cm}}{\textbf{Number of domiciles of source country spreading the 
infection}} & \multicolumn{1}{l}{\textbf{GIEP}} \\  \hline
China & 19 & 0.4043 \\
Italy & 14 & 0.2979 \\
United States of America & 5 & 0.1064 \\
United Kingdom & 4 & 0.0851 \\
Germany & 3 & 0.0638 \\
Iran & 3 & 0.0638 \\
Spain & 3 & 0.0638 \\
French Republic & 2 & 0.0426 \\
India & 2 & 0.0426 \\
Japan & 2 & 0.0426 \\
Netherlands & 2 & 0.0426 \\
Russian Federation & 2 & 0.0426 \\
Australia & 1 & 0.0213 \\
Belgium & 1 & 0.0213 \\
Democratic Republic of the Congo & 1 & 0.0213 \\
Ecuador & 1 & 0.0213 \\
Morocco & 1 & 0.0213 \\
Nigeria & 1 & 0.0213 \\
Norway & 1 & 0.0213 \\
Philippines & 1 & 0.0213 \\
Turkey & 1 & 0.0213 \\
United Arab Emirates & 1 & 0.0213 \\ \hline
\end{tabular}
\label{tab:ISP}
\end{table}

\begin{table}[htbp]
\caption{R adjusted ($R_{adj}$) and \textit{p-values} obtained by fitting 
different models on outdegree and density obtained from both C19-TraNet and 
corresponding SSF models (for comparison, average of $1,000$ SSF models is used.)}
\begin{tabular}{lrrrrrrrr}
\hline
\textbf{} & \multicolumn{ 4}{c}{\textbf{Outdegree}} & \multicolumn{ 
4}{c}{\textbf{Density}} \\
\textbf{} & \multicolumn{ 2}{c}{\textbf{Real}} & \multicolumn{ 
2}{c}{\textbf{SSF}} & \multicolumn{ 2}{c}{\textbf{Real}} & \multicolumn{ 
2}{c}{\textbf{SSF}} \\
\textbf{} & \multicolumn{1}{c}{$\mathbf{R_{adj}}$} & 
\multicolumn{1}{c}{\textbf{\textit{p-value}}} & 
\multicolumn{1}{c}{$\mathbf{R_{adj}}$} & 
\multicolumn{1}{c}{\textbf{\textit{p-value}}} & 
\multicolumn{1}{c}{$\mathbf{R_{adj}}$} & 
\multicolumn{1}{c}{\textbf{\textit{p-value}}} & 
\multicolumn{1}{c}{$\mathbf{R_{adj}}$} & 
\multicolumn{1}{c}{\textbf{\textit{p-value}}} \\ \hline
\textbf{Linear} & 0.623 & 2.80E-19 & 0.599 & 3.58E-18 & 0.441 & 3.54E-12 & 0.567 
& 9.09E-17 \\
\textbf{Quadratic} & 0.762 & 1.01E-24 & 0.906 & 2.24E-37 & 0.725 & 5.95E-19 & 
0.889 & 5.14E-34 \\
\textbf{Cubic} & 0.887 & 6.54E-35 & 0.987 & 8.93E-70 & 0.882 & 1.96E-29 & 0.984 
& 1.02E-64 \\
\textbf{Quartic} & 0.97 & 1.44E-55 & 0.995 & 2.14E-84 & 0.957 & 2.12E-44 & 0.994 
& 1.72E-81 \\
\textbf{Quintic} & 0.977 & 1.02E-59 & 0.995 & 1.19E-84 & 0.972 & 5.75E-51 & 
0.994 & 3.29E-81 \\
\textbf{Log} & 0.906 & 4.97E-44 & 0.911 & 5.16E-45 & 0.822 & 1.20E-32 & 0.896 & 
3.00E-42 \\
\textbf{Exponential} & 0.536 & 1.50E-15 & 0.54 & 1.11E-15 & 0.906 & 4.57E-44 & 
0.929 & 5.17E-49 \\ \hline
\end{tabular}
\label{tab:Trend}
\end{table}

\end{document}